\documentclass[aps,prl,nofootinbib,10pt,tightenlines,notitlepage, twocolumn,floatfix,superscriptaddress,showkeys]{revtex4-2}

\usepackage[utf8]{inputenc}          
\usepackage{graphicx}         
\usepackage[usenames,dvipsnames]{xcolor}
\usepackage{array,dcolumn,longtable} 
\usepackage{amsmath,amssymb,amsfonts,slashed} 
\usepackage{mathtools}          
\usepackage[linktocpage,breaklinks]{hyperref}
\usepackage{mathrsfs}
\usepackage{txfonts}
\usepackage{bm}
\usepackage{stmaryrd}
\usepackage{tensor}
\usepackage[utf8]{inputenc}

\usepackage{epsfig}
\usepackage{epstopdf}

\usepackage{cleveref}

\definecolor{mred}{RGB}{127,0,25}
\definecolor{mdgr}{RGB}{51,51,51}
\definecolor{mag}{RGB}{211, 54, 130}
\definecolor{verm}{RGB}{164, 25, 0}

\hypersetup{colorlinks=true,
            citecolor=NavyBlue,
            linkcolor=NavyBlue,
            urlcolor=NavyBlue}
\usepackage{siunitx}                                  
\DeclareSIUnit{\fm}{\femto\metre}                     

\newcommand{\scri}{\mathscr{I}}

\def\bfDelta{\mbox{\boldmath $\Delta$}}
\def\bfxi{\mbox{\boldmath $\xi$}}

\newcount\hh
\newcount\mm
\mm=\time
\hh=\time
\divide\hh by 60
\divide\mm by 60
\multiply\mm by 60
\mm=-\mm
\advance\mm by \time
\def\hhmm{\number\hh:\ifnum\mm<10{}0\fi\number\mm}

\def\be{\begin{equation}}
\def\ee{\end{equation}}

\def\bftheta{\mbox{\boldmath $\theta$}}

\begin{document}
\title{An order-unity correction to Hawking radiation}

\author{Eanna E. Flanagan}\email{eef3@cornell.edu}
\affiliation{Department of Physics, Cornell University, Ithaca, NY 14853}
\affiliation{Cornell Laboratory for Accelerator-based Sciences and Education (CLASSE), Cornell University, Ithaca, NY 14853}

\begin{abstract}

When a black hole first forms, the properties of the emitted
radiation as measured by observers near future null infinity are very
close to the 1974 prediction of Hawking.  However, deviations grow
with time, and become of order unity after a time $t \sim M_i^{7/3}$,
where $M_i$ is the initial mass in Planck units.  After an evaporation
time the corrections are large: the angular distribution of the emitted
radiation is no longer dominated by low multipoles, with an exponential
falloff at high multipoles.  Instead,
the radiation is redistributed as a power law spectrum over a broad
range of angular scales, all the way down to the scale $\Delta \theta \sim 1/M_i$,
beyond which there is exponential falloff.
This effect is is a quantum gravitational effect, whose
origin is the spreading of the wavefunction of the black hole's
center of mass location caused by the kicks of the individual outgoing
quanta, discovered by Page in 1980.
The modified angular distribution of the Hawking radiation has an
important consequence: the number of soft hair modes that can
effectively interact with outgoing Hawking quanta increases from the handful of
modes at low multipoles $l$, to a large number of modes, of order
$\sim M_i^2$.  We argue that this change unlocks the Hawking-Perry-Strominger
mechanism for purifying the Hawking radiation.

\end{abstract}

\maketitle

\vspace{0.1cm}
\noindent \textit{Introduction}~--~In the half century since its
discovery, the Hawking evaporation of black holes and its associated
conundrums have proved to be a fertile source of insights and progress
in quantum gravity, from black hole thermodynamics to holography to
links between quantum information and geometry
\cite{1976PhRvD..14.2460H,Harlow:2014yka,Marolf:2017jkr}.
At the same time, unresolved theoretical tensions have
led to repeated scrutiny of the robustness of Hawking's predictions.
An evaporating black hole is characterized by the small dimensionless
parameter $1/M$, where $M$ is the mass in Planck units,
and there are small corrections that are perturbative in $1/M$, as
well as smaller corrections nonperturbative in $1/M$.  Large
corrections however have been elusive.

There is a subtlety in classifying the size of corrections to Hawking
radiation, related to the fact
that the number of relevant field modes $N \sim M^2$ is large, and 
fractional corrections to expected
values may be small for certain classes of operators but large
for other operators.  Suppose we decompose the Hilbert space ${\cal H}$ of
radiation states at future null infinity $\scri^+$ as the product ${\cal H} = {\cal H}_n
\otimes {\cal H}_n^\prime$, where ${\cal H}_n$ is the Hilbert space of
a certain set of $n$ modes with $n \le N$.
Given a correction $\Delta \rho$ to the density matrix $\rho$ on ${\cal H}$,
we define
\be
\varepsilon_{{\cal H}_n} = \| \, {\rm tr}_{{\cal H}_n^\prime} \Delta \rho \,\|,
\ee
where $\| A \| = {\rm tr} \sqrt{ A^\dagger A}$.
which gives a measure of the size of the correction to the state when restricted
to ${\cal H}_n$.
There exist perturbations $\Delta \rho$
for which $\varepsilon_{{\cal H}_n}$ is small whenever $n \ll N$, but
for which $\varepsilon_{{\cal H}_n}$ is 
nevertheless of
order unity when $n \sim N$.
Such corrections have long been anticipated for Hawking radiation, since an order-unity
correction to an entanglement entropy is required\footnote{The fact
that $\| \, \Delta \rho \,\| \, = O(1)$ is required follows from the
identity  
$|S(\rho + \Delta \rho) - S(\rho)| \le \, \| \, \Delta \rho \,\| \, {\rm log} \, d + 1/e$,
where $S$ is von Neumann entropy and $d$ is the dimension of the Hilbert space \cite{2000PhRvA..61f4301N},
together with $S(\rho+\Delta \rho) = 0$ and $S(\rho) \sim {\rm log} \, d \sim M^2$.}
for unitarity of the
evaporation process \cite{Harlow:2014yka,Marolf:2017jkr}.
Indeed, recent calculations using Euclidean path integrals have explicitly
shown that the time evolution of the entanglement
entropy of the Hawking radiation and the black hole is
consistent with unitarity
\cite{Almheiri:2019psf,Penington:2019npb,Almheiri:2019hni,Almheiri:2019qdq,Penington:2019kki,Almheiri:2020cfm}. 
Hence there are corrections to Hawking radiation that are of
order unity, for operators that involve $\sim N$ modes, although the
new computational techniques do not yet allow computation of the
corrected state.

In this Letter we confine attention to operators that act on $n
\ll N$ modes, for which the general expectation has been that
corrections to Hawking radiation are small.  We show that there are
corrections at the level of individual modes that are of order unity,
arising from quantum gravitational effects in the deep infrared.
The mechanism is straightforward: secularly increasing fluctuations in
the center of mass location of 
the black hole cause a change in the angular distribution of the
radiation, with most of the power being redistributed to small angular
scales.  Although the modifications to the radiation
do not directly impact the issue of how unitarity of the  evaporation
is achieved, we will argue that there is an
important indirect effect.

In the remainder of the Letter, we first give a heuristic argument for
the effect, then give a detailed derivation, and conclude with
a discussion of some implications.  Throughout we use Planck units
with $G = \hbar = c =1$.

\vspace{0.1cm}
\noindent \textit{Redistribution of power to small angular scales:
  brief heuristic argument}~--~
As described by Page \cite{PhysRevLett.44.301}, the emission of
Hawking radiation causes 
the uncertainty in the black hole's center of mass to grow with time.
This growth is easy to understand: each outgoing
quantum carries off a momentum $\sim M^{-1}$ in a random direction,
and the resulting
perturbation to the velocity of the black hole is of order
$\sim M^{-2}$.  Over an evaporation time $\sim M^3$ this single kick yields
a displacement of the center-of-mass position of the black hole of
order $\sim M$.  Over the course of the evaporation process we have
$N \sim M^2$ such kicks that accumulate as a random walk, giving a total
net uncertainty in the position of the black hole of order $\sim \sqrt{N} M \sim M^2$.

\begin{figure}[h]
  \begin{center}
    \includegraphics[width=3in]{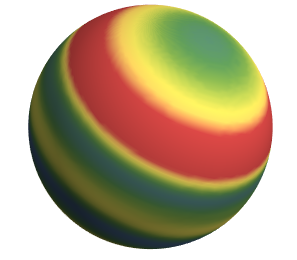}
    \caption{An illustration of the standard Unruh state of an evaporating non-spinning black hole,
      at a particular instant of retarded time at future null infinity, in a reference frame that is displaced from the black hole center of mass by several Schwarzschild radii.  The quantity plotted is a typical realization of the Gaussian random process
on the sphere whose two point function is given by taking the two
point function of a scalar field in the Unruh state at future null
infinity and subtracting the two point function of the out
vacuum. Fluctuations in individual wavepacket modes give rise to
fluctuations on the sphere that are confined to concentric thin
strips, giving rise to a characteristic angular scale that is small compared to unity.}
\label{fig-st}
    \end{center}
\end{figure}

Now if a black hole displaced by $\sim
M^2$ emits a single quantum in a wavepacket mode of duration $\sim M$,
the energy flux at future null infinity $\scri^+$ is delayed by $\sim M^2$ on
one side of $\scri^+$ and advanced on the other.  On a cut of fixed
retarded time, the energy flux due to this quantum will be localized
to a thin strip on the sphere of width $\Delta \theta \sim M/M^2 \sim 1/M$
(see Fig. \ref{fig-st}), and so the power
spectrum of the radiation as a function of angular scale will be
peaked at angular scales $\sim 1/M$.

\vspace{0.5cm}

\vspace{0.1cm}
\noindent \textit{Redistribution of power to small angular scales: derivation}~--~
Although the mechanism that modifies the Hawking radiation is
universal, for simplicity we will specialize here to a four dimensional
Schwarzschild black hole coupled to a massless free scalar field
$\Phi$. 
Near $\scri^+$ we use retarded Bondi coordinates
$(u,r,\theta,\phi)$.  We resolve the Bondi-Metzner-Sachs (BMS)
transformation freedom in these coordinates by choosing the canonical
coordinates\footnote{See, for example, Sec.\ II.C of Ref.\ \cite{FN}.}
associated with the approximate stationary state of the 
black hole shortly after it is formed at $u=0$ say, before it has time to emit
appreciable amounts of Hawking radiation\footnote{Note that imposing
this requirement in the region of $\scri^+$ near $u=0$
determines the coordinates everywhere on $\scri^+$ because of the
properties of the BMS group.}.  This choice also determines 
a particular Poincar\'e subgroup of the BMS group.

We define the field $\varphi$ on $\scri^+$ by
\be
   \Phi(u,r,\theta,\phi) = \frac{\varphi(u,\theta,\phi)}{r} + O \left( \frac{1}{r^2} \right),
\ee
and we denote by ${\cal H}$ the Hilbert space of out states on
$\scri^+$ parameterized by $\varphi$.
We denote by $M_i$ the initial mass of the black hole at $u=0$, and by
$M=M(u_1)<M_i$ the Bondi mass at some later retarded time $u_1$ with
$u_1  \gg M_i$.
We denote by $\rho_U$ the Hawking radiation state on ${\cal H}$ for
an eternal black hole of mass $M$, i.e. the Unruh vacuum.
In the standard calculation it is argued that this state
provides a good approximation to the $n$-point functions on $\scri^+$ of the state
for the
gravitational collapse spacetime, for retarded times $u$
with $|u - u_1|$ small compared to the evaporation time $M^3$.

For any density matrix $\rho$ on ${\cal H}$ we define the regularized two-point function
\begin{eqnarray}
  G(u,\bftheta; u',\bftheta')&=& {\rm tr} \left[ \rho \varphi(u,\bftheta) \varphi(u',\bftheta') \right] \nonumber \\
&&  - {}_{\rm out}\left<0 \right| 
  \varphi(u,\bftheta) \varphi(u',\bftheta') \left| 0 \right>_{\rm out},
  \label{twopt}
\end{eqnarray}
where $\left| 0 \right>_{\rm out}$ is the out vacuum and $\bftheta =
(\theta,\phi)$.  For stationary, spherically symmetric states we have
$G = G( \Delta u,\gamma)$, where $\Delta u = u - u'$ and $\gamma$ is the angle between
$\bftheta$ and $\bftheta'$.
We define the Fourier transform ${\tilde G}(\omega,\gamma) = \int
d\Delta u \, e^{i \omega \Delta u} \, G(\Delta u,\gamma)$, and
decompose this in angular harmonics as
\be
{\tilde G}(\omega, \gamma) =  \sum_{l=0}^\infty  \frac{ 2 l + 1}{4 \pi} P_l(\cos \gamma)
S(\omega,l).
\label{Sdef}
\ee
The quantity $S(\omega,l)$ is related to the energy flux
${\dot E}$ to infinity per unit frequency $\omega$ in field
multipoles\footnote{Note that there are two different methods of
defining the angular spectrum of Hawking radiation.  One can decompose
the field into spherical harmonic modes, or decompose the stress
energy tensor.  We use the former definition.  The two definitions are
not equivalent since the stress energy tensor depends nonlinearly on
the field.  Nevertheless, the qualitative result that power is
redistributed from large angular scales to small angular scales will clearly
be be valid for both definitions.}
of order $l$ by
\be
\left( \frac{d {\dot E}}{d \omega} \right)_l = \frac{2 l + 1}{2 \pi}
\omega^2 S(\omega, l).
\label{Sdef1}
\ee
We denote by $G_U$ the regularized two point function of the Unruh vacuum, for
which the corresponding energy
flux is
\begin{eqnarray}
  \left( \frac{d {\dot E}}{d \omega} \right)_{{\rm U},l} =
  \frac{2l+1}{2 \pi}  \frac{
    \omega |t_{l\omega}|^2 }{e^{\beta \omega} -1}.
  \label{standard}
\end{eqnarray}
Here $t_{l\omega}$ is the transmission coefficient through the
effective potential and $\beta = 8 \pi M$ is the inverse temperature
of the radiation.  As is well known, most of the power in the spectrum (\ref{standard})
is concentrated at $l \sim O(1)$, with an exponential falloff at large $l$.

We now want to derive how the energy spectrum (\ref{standard}) as a
function of frequency and angular scale is modified.
The key idea is to supplement the standard computation by including
the evolution of a small number of relevant infrared gravitational
degrees of freedom, specifically the BMS charges as computed on cuts
$u = $ const of $\scri^+$.  
In the classical theory, the values of these charges determine the
spacetime geometry when the black hole is stationary, and we assume
that this is still true in the quantum theory when both the charges
and geometry have quantum fluctuations.
We focus in
particular on the black hole's center of mass $\bfDelta$, encoded in the orbital
angular momentum associated with the Poincar\'e subgroup of the BMS
group discussed above.  

The framework we use is anchored at $\scri^+$, where the coordinate system $(u,\theta,\varphi)$ and
out Hilbert space ${\cal H}$ for the scalar field are unaffected by the
large quantum fluctuations of the gravitational charges and of the geometry in the interior.
As described above, the fluctuations in $\bfDelta$ grow with time due
to repeated kicks from outgoing Hawking quanta.
We divide $\scri^+$  
into an early portion $\scri^+_{\rm early}$ with $u<u_1$, and a late
portion $\scri^+_{\rm late}$ with $u > u_1$.  The Hilbert space ${\cal H}$ can be correspondingly 
factored\footnote{This is not quite true, as there
are also ``edge modes'' associated with the boundary $u=u_1$
\cite{Donnelly:2016auv,Speranza:2017gxd,2017NuPhB.924..312G},
equivalent to soft hair \cite{Hawking:2016msc,Hawking:2016sgy}.  We
neglect these modes here as they not relevant to the present 
discussion, but will return to them later in the paper.} into the
tensor product ${\cal H}_{\rm early} \otimes {\cal 
  H}_{\rm late}$.  The state of the center-of-mass at time $u_1$ is
strongly correlated with the Hawking radiation on $\scri^+_{\rm
  early}$, by momentum conservation for each emission event, and if we trace over ${\cal H}_{\rm early}$ we 
obtain a mixed state for the center-of-mass\footnote{One can think of this as the early Hawking radiation
decohering the black hole location \cite{Arrasmith:2017ogi}.}. This state can be
described in terms of its Wigner function ${\cal W}({\bf \Delta}, {\bf p})$, a function of
the three dimensional position ${\bf \Delta}$ and momentum ${\bf p}$ of the
black hole.  Denoting a position eigenstate by $\left| \bfDelta
\right>$, the corresponding state is
\be
\int d^3 \Delta \int d^3 \xi \, {\widetilde {\cal W}}(\bfDelta,\bfxi) \, \left|
\bfDelta - \bfxi/2 \right> \left< \bfDelta + \bfxi/2 \right|,
\label{comstate}
\ee
where ${\widetilde {\cal W}}(\bfDelta,\bfxi) = \int d^3p \exp[- i {\bf p}
  \cdot \bfxi] {\cal W}(\bfDelta, {\bf p})$.
Since the kicks from the individual outgoing quanta are
uncorrelated, the Wigner function ${\cal W}$ is very nearly Gaussian by the
multivariate central limit theorem.  Hence ${\widetilde {\cal W}}$ has the form
\begin{eqnarray}
{\widetilde  {\cal W}}(\bfDelta,\bfxi) =
{\cal N} \exp \left[ - \frac{1}{2}
   \frac{ \bfDelta^2}{\sigma_\Delta^2} - \frac{1}{2}   (1 - \varepsilon^2)
  \bfxi^2 \sigma_p^2 - i \varepsilon \frac{\sigma_p}{\sigma_\Delta} \bfDelta \cdot
  \bfxi \right],
\label{wigner}
\end{eqnarray}
where ${\cal N} =   (2 \pi)^{-3/2}
\sigma_\Delta^{-3}$, the quantities $\sigma_\Delta^2$ and $\sigma_p^2$ are the
variances in position and momentum, and $\varepsilon$ with
$|\varepsilon|<1$ is a correlation coefficient.
The evolution of these parameters is studied in 
Ref.\ \cite{companion}, which shows that $\varepsilon$ is of order
unity and
\FL
\begin{eqnarray}
\label{params}
\sigma_\Delta^2 &=& \left\{ \begin{array}{ll} 
         c_0 M_i^4 (1 - M^3 /M_i^{3})^3   & \mbox{ $M_i-M \ll M_i$,} \nonumber \\
         c_1 M_i^4
          & \mbox{    $\sqrt{M_i} \ll M \ll M_i$}, \\
		\end{array}
\right. \\
\sigma_p^2 &=& c_2 \ln( M_i / M) \ \ \ \ \ \ \ \ \ \ \ \ \ \ \ \ \ \  \mbox{$\sqrt{M_i} \ll M$},
\end{eqnarray}
where $c_0$, $c_1$ and $c_2$ are dimensionless constants of order unity.

We now turn to describing how the fluctuations in the center of mass of the black hole affect the Hawking radiation.
In Minkowski spacetime we can define a displacement operator  $U_{\Delta}$
which displaces any state by an amount $\bfDelta$,
which acts on the field operator according to $U_{\Delta}^\dagger \Phi(t,{\bf r}) U_{\Delta} = \Phi(t, {\bf r} - \bfDelta)$.
This operator extends naturally to the Hilbert space ${\cal H}$ of out states on the black hole spacetime,
where its action is defined by
\be
U_{\Delta}^\dagger \varphi(u,\bftheta) U_{\Delta} = \varphi(u + {\bf
  n} \cdot \bfDelta, \bftheta)
\label{Uaction}
\ee
with ${\bf n}$ the unit vector in the direction specified by
$\bftheta$.  The Unruh state for a black hole displaced from the
origin by an amount $\bfDelta$
can be written as
\be
\left| \bfDelta \right> \otimes \sum_j \left| \chi_j \right> U_\Delta \left| \psi_j \right>,
\label{displacedU0}
\ee
where $\left| \chi_j \right>$ is a set of states on the future horizon and $\left| \psi_j \right>$ a set of states
in ${\cal H}$.  Taking the trace over the horizon states gives for the corresponding Unruh state at $\scri^+$
\be
\left| \bfDelta \right>  \left< \bfDelta \right| \otimes U_\Delta
\rho_{\rm U} U_\Delta^\dagger,
\label{displacedU}
\ee
where $\rho_U = \sum_j c_j^2 | \psi_j> < \psi_j |$
with $c_j^2 = < \chi_j | \chi_j >$.

Suppose now that the state of the black hole's center of mass were
fixed and not evolving with time, given by Eq.\ (\ref{comstate}) for the
fixed values of the parameters $\sigma_\Delta$, $\sigma_p$ and
$\varepsilon$ evaluated at $u=u_1$.  Then by linearity
from
Eqs.\ (\ref{comstate}), (\ref{displacedU0}) and (\ref{displacedU}) the corresponding out state would be
\be
\int d^3 \Delta \int d^3 \xi \, {\widetilde {\cal W}}(\bfDelta,\bfxi)
\,
\left| \bfDelta -\bfxi/2\right>  \left< \bfDelta +\bfxi/2\right| \otimes 
U_{\Delta-\xi/2} \, \rho_{\rm U} \, U_{\Delta+\xi/2}^\dagger.
\label{Ucorr0}
\ee
Tracing over the center of mass Hilbert space gives the
corrected version of the Unruh state
\be
\rho_{\rm{U,corr}} = 
\int d^3 \Delta \, {\widetilde {\cal W}}(\bfDelta,{\bf 0})
\,
U_{\Delta} \, \rho_{\rm U} \, U_{\Delta}^\dagger.
\label{Ucorr}
\ee

Of course the state of the center of mass is evolving with time and
not fixed.  Nevertheless, the corrected Unruh state (\ref{Ucorr})
should give a good approximation to the $n$-point functions on
$\scri^+$ of the field at retarded times $u$ that satisfy two
conditions:
\begin{itemize}
\item We have $|u - u_1| \ll M^3$, so the mass of the black hole as
  well as the state of the center-of-mass have not evolved
  significantly from their values at $u = u_1$.

  \item We have $u - u_1 \gg \sigma_\Delta \sim M_i^2$.  This ensures
    that the displacements (\ref{Uaction}) in retarded time caused by the operators
    $U_\Delta$ in Eq.\ (\ref{Ucorr}) do not generate a dependence on
    degrees of freedom on $\scri^+_{\rm early}$, which we have
    already traced over to compute the state (\ref{comstate}).
\end{itemize}

We now turn to showing that the modifications inherent in the
corrected Unruh state (\ref{Ucorr}) are of order unity, for individual
outgoing wavepacket modes at sufficiently late times.  Combining Eqs.\ (\ref{twopt}), (\ref{Uaction}) and
(\ref{Ucorr}) gives for the regularized two point function of the corrected
Unruh state
\begin{eqnarray}
G_{\rm U,corr}(u,\bftheta;u',\bftheta') &=&
\int d^3 \Delta \, {\widetilde {\cal W}}(\bfDelta,{\bf 0})
\nonumber \\ && \times
G_{\rm U}(u + {\bf n} \cdot
\bfDelta, \bftheta ; u' + {\bf n}' \cdot \bfDelta, \bftheta'),
\ \ \ \ \ 
\end{eqnarray}
using that the Wightman function in the second term in Eq.\ (\ref{twopt}) is invariant under
translations.  The corresponding functions of frequency $\omega$ and
angle $\gamma$ are related by 
\begin{eqnarray}
{\tilde G}_{\rm U,corr}(\omega,\gamma) &=&
\int d^3 \Delta \, {\widetilde {\cal W}}(\bfDelta,{\bf 0})
e^{ - i \omega ({\bf n} - {\bf n}')\cdot \bfDelta}
{\tilde G}_{\rm U}(\omega, \gamma) \nonumber \\
&=& \exp \left[ - 2 \omega^2 \sigma_\Delta^2 \sin^2(\gamma/2) \right]
{\tilde G}_{\rm U}(\omega,\gamma),
\label{transform}
\end{eqnarray}
where we have used Eq.\ (\ref{wigner}).
Note that the transformation (\ref{transform}) preserves ${\tilde
  G}(\omega,0)$ which is proportional to the total energy flux per
unit frequency, summed over all multipoles.  Hence the transformation
redistributes power over angular scales, but not from one
frequency to another.

We next combine Eqs.\ (\ref{Sdef}), (\ref{Sdef1}) and
(\ref{transform}) to obtain for the spectrum of outgoing
radiation
\begin{eqnarray}
  \left( \frac{d {\dot E}}{d \omega} \right)_{{\rm U, corr},l} =
  (2l+1) \omega^2 \int_{-1}^1 d \mu P_l(\mu) e^{- \omega^2
    \sigma_\Delta^2 (1-\mu)} 
{\tilde G}_{\rm U}(\omega,\gamma), \ \ \ \ \ 
  \label{newspectrum}
\end{eqnarray}
where $\mu = \cos\gamma$.
We now specialize to frequencies of the order $\omega \sim M^{-1}$,
where most of the outgoing power is located.  We thus exclude
high frequencies $\omega \gg M^{-1}$ where the power is exponentially
suppressed, and low frequencies $\omega \ll M^{-1}$ where it is
power-law suppressed, from the spectrum (\ref{standard}).   
Since the function $\omega {\tilde G}_{\rm U}(\omega,\gamma)$ depends on
$\omega$ and $M$ only through the combination $\omega M$
\cite{Gray:2015xig}, which is of
order unity, for such
frequencies ${\tilde G}_{\rm U}$ varies with $\gamma$ only on angular scales
of order unity; there are no other dimensionless parameters on
which the function depends. It follows that 
${\tilde G}_{\rm U}$ 
has negligible variation over the range $0 \le \gamma \alt 1/(\omega \sigma_\Delta) \sim
M/M_i^2 \ll 1$ that is not exponentially suppressed by the exponential
factor in Eq.\ (\ref{newspectrum}).  Hence we can evaluate this
function at $\gamma=0$ and pull it outside the integral, and using
Eqs.\ (\ref{Sdef}) and (\ref{Sdef1}) we re-express it in terms of the total
power per unit frequency $d {\dot E} /d \omega = \sum_l ( d {\dot E}/d
\omega)_l$ in the Unruh state.  We evaluate the remaining integral
using the identity $\int d\mu P_l(\mu) e^{a \mu} = \sqrt{2 \pi/a}
I_{l + 1/2}(a)$ which expresses it terms of a modified Bessel
function of the first kind \cite{10.5555/1830479}.  The final result
is
\begin{eqnarray}
  \left( \frac{d {\dot E}}{d \omega} \right)_{{\rm U, corr},l} =
\sqrt{\frac{\pi}{2}}   \left( \frac{d {\dot E}}{d \omega} \right)_{{\rm U}} 
\frac{(2l+1) e^{-\omega^2 \sigma_\Delta^2}}{\omega \sigma_\Delta}
I_{l + 1/2}( \omega^2 \sigma_\Delta^2). \ \ \ \ \ 
  \label{newspectrum1}
\end{eqnarray}
Using the approximate formula $I_{l+1/2}(a) = (2 \pi a)^{-1/2} e^a [ 1
  + O(l^2/a) ]$ this simplifies to\footnote{For $l \gg 1$
Eq.\ (\ref{newspectrum2}) can be more simply derived by approximating
the sphere as a plane and replacing the transform (\ref{newspectrum}) with a two
dimensional Fourier transform.}
\begin{eqnarray}
  \left( \frac{d {\dot E}}{d \omega} \right)_{{\rm U, corr},l} =
   \left( \frac{d {\dot E}}{d \omega} \right)_{{\rm U}} 
\frac{(2l+1)}{2 \omega^2 \sigma^2_\Delta}
\left[ 1 + O\left( \frac{l^2}{\omega^2 \sigma_\Delta^2} \right) \right]. \ \ \ \ \ 
  \label{newspectrum2}
\end{eqnarray}
This corresponds to a power-law spectrum for angular scales in the
range $0 \le l \ll l_{\rm crit}$ with $l_{\rm crit} = \omega
\sigma_\Delta$, with most of the power in the
vicinity of $l \sim l_{\rm crit}$.  At scales $l \ge l_{\rm crit}$
the spectrum falls off exponentially, from the upper bound
$I_{l+1/2}(a) \le (2 \pi a)^{-1/2} e^a \exp [ - l/ (4 \sqrt{a})]$ for $l \ge \sqrt{a} \gg 1$.

We now consider the critical angular scale $l_{\rm crit} = \omega
\sigma_\Delta$.  At sufficiently late times $u \agt M_i^3$ we have $\sigma_\Delta \sim M_i^2$ from
Eq.\ (\ref{params}), and so the critical angular scale is $l_{\rm
  crit} \sim M_i^2/M \gg 1$ using $\omega \sim M^{-1}$, which reduces to $\sim M_i$ if $M \sim M_i$.
At early times we have from Eq.\ (\ref{params}) and using 
$u/M_i^3 \sim 1 - M^3/M_i^3$ that $l_{\rm crit} \sim u^{3/2}
M_i^{-7/2}$, so the modification effect first becomes of order unity
after an interval of retarded time $u \sim M_i^{7/3}$.

\vspace{0.1cm}
\noindent \textit{Discussion and conclusions:}~--~We close with a
number of comments. First, the modification to
the Hawking radiation does not alter the amount
of entanglement between modes inside the horizon and those outside,
and so does not directly impact the unitarity of the evaporation
process.  The exterior modes that are relevant at late times depend,
through the position of the black hole,
on which early time exterior modes are occupied (the total number of relevant
exterior modes has increased from $\sim M_i^2$ to $\sim M_i^4$).
This effect generates nontrivial mutual information \cite{Harlow:2014yka}
between early Hawking radiation and late Hawking radiation, but does
not alter the total entanglement between interior and exterior modes.

Second, the corrected Unruh state (\ref{Ucorr}) is
not a Gaussian state, unlike the original Unruh state, although it is
stationary and spherically symmetric.  Thus it is not determined by
the two-point function (\ref{newspectrum1}), although it is completely
determined by the formulae (\ref{wigner}) and (\ref{Ucorr}).

Third, the general mechanism 
discussed here involving spatial translations clearly also applies to
other generators of the BMS group.   The black hole at late
times determines a BMS frame which is related to the initial BMS frame
by a transformation which includes a rotation, boost and
supertranslation, and secularly growing quantum fluctuations in
those transformations modify the outgoing Hawking radiation.
However, in Ref.\ \cite{companion} we estimate that the typical
boost\footnote{The effect of boosts is also suppressed by the fact
that the operator describing the evolution of the quantum field is
approximately diagonal on a position basis for the black hole center
of mass, but not on a momentum basis.  This is why boost
fluctuations are not present in Eq.\ (\ref{Ucorr}), although they are
present in the Wigner function ${\cal W}$.}
velocity scale is $\sim 1/M$, and that the lengthscale involved in the
supertranslation fluctuations is $\sim 1$, so the corresponding
modifications to the Hawking radiation are small.  

Fourth, consider the result of interpreting the corrected Unruh state (\ref{Ucorr}) on $\scri^+$ in
terms of a single semiclassical spacetime with the black hole at the origin.
An outgoing mode with $l \sim M_i$ near $\scri^+$ corresponds near the black hole to
a large amplitude standing wave in a thin shell of width $\sim 1$
in the non-evanescent region between the horizon and potential barrier, which varies over transverse lengthscales along the horizon
of order $\sim 1$.  This Planckian behavior of the extrapolated corrected Unruh state 
illustrates the potential pitfalls of thinking in terms of a single semiclassical
spacetime and focusing on near-horizon physics.


Fifth, we argue that the modification to the Hawking process removes
one of the primary objections to the proposal that soft hair on
black holes plays a key role in resolving the information loss paradox
\cite{Hawking:2016msc,Hawking:2016sgy,Strominger:2017aeh,Pasterski:2020xvn,Cheng:2020vzw}.
Soft hair consists of charges measurable at future null infinity
associated with an extension of the BMS group \cite{Strominger:2017zoo,Campiglia:2014yka,CL,Cnew},
higher-$l$ analogs of the center-of-mass that are
encoded in the asymptotic metric.  Just as for the center-of-mass, the expected value of soft hair charges
can be set to zero by a gauge transformation, locally in time, but their variances
cannot and can contain nontrivial information.  Outgoing Hawking
quanta excite soft hair via the gravitational wave memory effect.
It has been suggested that the Hawking radiation is purified at late
times by its entanglement with soft hair degrees of freedom
\cite{Strominger:2017aeh}.

A difficulty with this proposal has been that only low $l$ modes of
the soft hair can be excited by the outgoing quanta, because of the
exponential falloff of the spectrum (\ref{standard}) at high $l$.
The soft hair field $\Phi(\bftheta)$ is given in terms of the scalar
field $\varphi$ on $\scri^+$ by [see, eg. Eqs.\ (2.19) and (4.4)
  of Ref.\ \cite{FN}]
\be
D^2 (D^2 + 2) \Phi(\bftheta) = 32 \pi {\cal P} \int du \,
\varphi_{,u}(u,\bftheta)^2,
\label{memory}
\ee
where $D^2$ is the Laplacian on the two-sphere and ${\cal P}$ is a
projection operator that sets to zero $l=0,1$ modes.  Thus only a
handful of soft hair modes can be excited, to few to play a relevant role for purifying the $\sim M_i^2$ outgoing Hawking quanta.

The modified angular distribution of the Hawking radiation completely
changes this picture, since the source term in Eq.\ (\ref{memory})
now extends effectively up to multipoles of order $l \sim M_i$. This makes
$\sim M_i^2$ soft hair modes potentially accessible, enough for each
outgoing quantum to interact with its own soft hair mode.  Note however
that this scenario cannot be analyzed within a single semiclassical
spacetime.  The details of the interaction of the Hawking radiation
with the soft hair is an intriguing subject for further study.

\vspace{0.1cm}
\noindent \textit{Acknowledgments}~--~ I thank Abhay Ashtekar,
Venkatesa Chandrasekaran and Kartik Prabhu for helpful discussions,
and an anonymous referee for useful comments. 
This research was supported in part by NSF grants PHY-1404105 and PHY-1707800.

\bibliography{../infoloss}

\end{document}